\documentstyle[preprint,aps]{revtex}

\begin{document}
\topmargin -1.4cm

\draft
\title{Hamiltonian approach to the ac Josephson effect in 
superconducting-normal hybrid systems}

\author{Qing-feng Sun and Hong Guo}

\address{{Center for the Physics of Materials and Department
of Physics, McGill University, Montreal, PQ, Canada H3A 2T8.}}

\author{Jian Wang}

\address{{Department of Physics, University of Hong Kong, 
Pokfulam Road, Hong Kong, China.}}


\maketitle

\begin{abstract}
The ac Josephson effect in hybrid systems of a normal mesoscopic conductor 
coupled to two superconducting (S) leads is investigated theoretically. 
A general formula of the ac components of time-dependent current is 
derived which is valid for arbitrary interactions in the normal region. 
We apply this formula to analyze a S-normal-S system where the normal
region is a noninteracting single level quantum dot. We report the physical
behavior of time-averaged nonequilibrium distribution of electrons in the 
quantum dot, the formation of  Andreev bound states, and ac components of
the time-dependent current. The distribution is found to exhibit a population 
inversion; and all Andreev bound states between the superconducting gap 
$\Delta$ carry the same amount of current and in the same flow direction.
The ac components of time-dependent current show strong oscillatory 
behavior in marked contrast to the subharmonic gap structure of the 
average current.
\end{abstract}



\section{INTRODUCTION}

Quantum transport properties of mesoscopic conductors coupled to 
two superconducting (S) leads have been extensively investigated
in the last decade both theoretically and 
experimentally\cite{ref1,ref2}. The mesoscopic conductor in question
is usually not a superconductor itself, but it can
be a quantum point contact (QPC)\cite{ref3,ref4,ref5}, 
a quantum dot (QD)\cite{ref6,ref7,ref8,ref9}, a tunnel barrier, 
a normal metal\cite{ref10,ref11}, and even a molecule such 
as a nanotube\cite{ref14,ref15,ref16}. The physics of these hybrid
device structures, in the form of S-normal-S, has profound 
implications to both fundamental understanding of quantum transport 
at reduced dimensionality and to practical applications in 
nanoelectronics. 

One of the main transport characteristics of a S-normal-S device 
structure is that particles in the normal region can undergo multiple 
Andreev reflections by the two superconducting leads. 
If the normal region is ballistic, a consequence of the
coherent superposition of these multiple Andreev reflections
is the formation of Andreev bound states\cite{ref1,ref17}. 
The Andreev bound states are important because they carry
current including the supercurrent. On the other hand, if 
the normal region is diffusive, a so-called supercurrent-carrying 
density of states, instead of the Andreev bound states, gives
the ability for carrying supercurrent\cite{ref18}. The
multiple Andreev reflection is also known to generate
subharmonic gap structure in the behavior of $I_0=I_0(V)$,
where $I_0$ is the average current and $V$ is the bias 
voltage\cite{ref3,ref4,ref5,ref6,ref7,ref8,ref10}. More recently, 
the subharmonic gap structure is used to measure transmission 
probability of each channel in a multi-channel QPC 
device\cite{ref19,ref20}. 

Another important and interesting transport characteristic of 
S-normal-S devices is the Josephson effect which gives rise to a 
dc supercurrent at zero bias, and an ac current at non-zero bias.
Previous theoretical analysis have focused on the dc Josephson effect
at zero bias\cite{ref9}, and the subharmonic gap structure of the 
{\it average} current at a non-zero 
bias\cite{ref3,ref4,ref5,ref6,ref7,ref8,ref10}. However, the ac 
Josephson effect, which arises at a non-zero bias, produces a 
current that is a function of time $t$. Therefore it is an important
task to theoretically understand the time dependent current in 
addition to understanding its time-average. To the best of our 
knowledge, so far there have been only two works which involve a 
time-dependent current\cite{ref4,ref5} for S-normal-S devices. 
Cuevas {\sl et. al.} have investigated the ac component of the 
time-dependent current for a S-QPC-S 
system\cite{ref4}. Bratus {\sl et. al.}\cite{ref5} have
investigated the time-dependent current in a
S-quantum-constriction-S system by considering an arbitrary normal 
electron transparency and discussing the current property at the
small bias limit. In both these works, the normal region of the 
S-normal-S device was simplified so as not to have any electronic 
structure: it is simply described by a constant transmission
coefficient which is independent of energy $\epsilon$. 
Given the interesting physics already discovered by these previous
investigations, it is indeed not difficult to expect that even richer
physics would arise if the normal region has its own electronic
structure. 

It is the purpose of this work to further investigate the ac
Josephson effect in S-normal-S device systems, and we focus on
issues not resolved by the previous analysis. In particular, we
consider a mesoscopic S-normal-S device with an arbitrary normal
region which may have its own electronic structure and/or
strong electron-electron interactions: for this general situation we
have derived the expression of the ac current. As an application we
then investigate a specific case in which the normal region 
is a ballistic quantum dot having a noninteracting single energy 
level, for which we investigate the intradot nonequilibrium 
distribution of electrons, the local density of state (LDOS), 
and the time-dependent current. Our main findings are: (i) The 
intradot electronic distribution shows a population inversion 
property. This property is distinctly and qualitatively 
different from that of the case where the normal region is 
diffusive. (ii) At small bias voltages such that $eV < \Delta$ where
$\Delta$ is the superconducting gap, a series of Andreev quasi-bound 
states is found to emerge within the gap. Their weights are not 
the same but they carry equal amount of current in the same direction, 
as well, their electronic occupations are all 1/2. 
This is qualitatively different from that of the zero bias case in
which the successive Andreev bound states carry opposite current.
(iii) The ac current component versus bias $V$ shows an 
oscillatory behavior. The amplitude of oscillation of the $n$th 
component is largest at about $V=\Delta/n$. At small bias, the 
high-order components quickly increase, and the time-dependent
current versus time $t$ deviates from a sine-like curve.

The rest of the paper is organized as follows. In Sec. II, 
the model Hamiltonian is presented and a general formula for the 
ac current component is derived. In Sec. III, ac Josephson effect 
for a simple S-normal-S device with a noninteracting normal region
is investigated. The intradot electronic distributions, the Andreev
quasi-bound states, and ac current components 
are presented in this Section. Finally, a brief summary is given in Sec. IV.

\section{Model and formulation} 

We assume the S-normal-S device system to be described by 
the following Hamiltonian:\cite{ref21,ref22}
\begin{equation}
H=\sum\limits_{\alpha=L,R} H_{\alpha} + H_{cen}+ H_T \hspace{1mm},
\label{hh1}
\end{equation}
where
\begin{eqnarray}
& &
H_{\alpha}= \sum\limits_{k,\sigma} \epsilon_{\alpha k} 
a^{\dagger}_{\alpha k \sigma} a_{\alpha k \sigma}
+\sum\limits_k\left[ \Delta_{\alpha} 
a_{\alpha k \downarrow} a_{\alpha -k\uparrow}
+\Delta_{\alpha} a^{\dagger}_{\alpha -k\uparrow} 
a^{\dagger}_{\alpha k \downarrow}
\right]\hspace{1mm}, \\
& &
H_{cen}= \sum\limits_{j,\sigma} \epsilon_{j\sigma} 
c^{\dagger}_{j\sigma} c_{j\sigma}
+H_{int}(\{c^{\dagger}_{j\sigma}\},\{c_{j\sigma}\},...)\hspace{1mm},\\
& &
H_T= \sum\limits_{k,j,\sigma,\alpha} t_{\alpha j}
e^{\frac{i}{2}\left(\phi_{\alpha} +
\frac{2eV_{\alpha}}{\hbar}t\right)}
a^{\dagger}_{\alpha k\sigma} c_{j\sigma} + H.c.
\end{eqnarray}
$H_{\alpha}$ ($\alpha=L,R$) describes the left/right 
BCS superconducting lead with the superconducting energy gap 
$\Delta_{\alpha}$. $H_{cen}$ is the Hamiltonian of the normal 
region of the device, and $c^{\dagger}_{j\sigma}(c_{j\sigma})$ 
are the creation (annihilation) operators
of an electron in state $j\sigma$ of the normal region. 
$H_{int}$ models interactions in the normal region whose
form depends on specific physics problems under consideration.
In this Section we consider the general case without 
specifying its concrete form. In deriving the formula for the
transport current, we permit the device normal (central) region 
to have various interactions, such as the electron-electron 
Coulomb interaction,
$\sum\limits_{j,\sigma; j_1,\sigma_1(j\sigma\not=j_1\sigma_1)} 
U_{j\sigma;j_1\sigma_1} c^{\dagger}_{j\sigma}c_{j\sigma}
c^{\dagger}_{j_1\sigma_1}c_{j_1\sigma_1}$; 
the electron-phonon interaction, 
$\sum\limits_{j,\sigma,q} M_{jq} c^{\dagger}_{j\sigma}c_{j\sigma} 
(d^{\dagger}_q +d_{-q}) +\sum\limits_q \hbar\omega_q 
d^{\dagger}_q d_q $; the tunneling coupling between different 
states of the normal region, $\sum\limits_{i,j,\sigma(i\not=j)}
\left[t_{ij} c^{\dagger}_{i\sigma}c_{j\sigma} +H.c.\right]$; 
and so on. $H_T$ of Eq.(\ref{hh1}) denotes the tunneling 
Hamiltonian between the superconducting leads and the normal 
region of the device, and $t_{\alpha j}$ is the hopping
matrix. In order to obtain the Hamiltonian (1), we have 
performed a unitary transformation, then the superconducting 
initial phase $\phi_{\alpha}$ and the terminal voltage 
$V_{\alpha}$ emerge in the Hamiltonian $H_T$.\cite{ref21,ref22}

The total current of superconducting lead $\alpha$
({\it e.g.} $\alpha=L$) flowing into the device normal 
region can be calculated from evolution of the total
number operator of electrons in that lead,
$N_L=\sum_{k,\sigma} a^{\dagger}_{Lk\sigma} a_{Lk\sigma}$. 
Then we have (in units of $\hbar=1$):\cite{ref22,ref23,ref24}
$$ I_L(t) =-e<\dot{N}_L(t)>=ie <[N_L,H]> \hspace{35mm}$$
\begin{equation}
=2e Re\sum\limits_{k,i} t_{Li} e^{i\left(\frac{\phi_L}{2}
+eV_L t\right)}
Tr\left\{ \hat{G}^<_{i,Lk}(t,t) \hat{\sigma}_z\right\}
\hspace{1mm},
\end{equation}
where
$$\hat{G}^<_{i,Lk}(t,t_1)\equiv i
\left( \begin{array}{ll}
<a^{\dagger}_{Lk\uparrow}(t_1) c_{i\uparrow}(t)> &
<a_{L-k\downarrow}(t_1) c_{i\uparrow}(t)>   \\
<a^{\dagger}_{Lk\uparrow}(t_1) c^{\dagger}_{i\downarrow}(t)> &        
<a_{L-k\downarrow}(t_1) c^{\dagger}_{i\downarrow}(t)>   \\ 
\end{array}
\right)
$$ 
is the distribution Green's function in the $2\times 2$ Nambu 
representation, and $\hat{\sigma}_z$ is the Pauli matrix. 
In this paper, we use the notation that ``$\hat{A}$" means 
quantity $A$ to be a $2\times 2$ matrix.

To proceed we need to solve the Green's function $\hat{G}^<_{i,Lk}(t,t)$. 
We assume that the leads do not have any interactions except the quadratic 
pair potential correlation, we have:\cite{ref22,ref24}
\begin{equation}
\hat{G}^<_{i,Lk}(t,t) = \sum\limits_j \int dt_1 \left[
\hat{G}^r_{ij}(t,t_1)\hat{t}^*_{Lj}(t_1)\hat{g}^<_{Lk}(t_1,t) +
\hat{G}^<_{ij}(t,t_1)\hat{t}^*_{Lj}(t_1)\hat{g}^a_{Lk}(t_1,t)
\right] \hspace{1mm},
\end{equation}
where $\hat{g}^{<,a}_{Lk}(t_1,t)$ is the exact Green's function 
of the left superconducting lead,\cite{ref4,ref21} and 
$\hat{t}^*_{Lj}(t_1)$ is a $2\times 2$ hopping matrix defined by:
\begin{equation}
\hat{t}^*_{Lj}(t)=\left( \begin{array}{ll}
t_{Lj}e^{i\left(\frac{\phi_L}{2}+eV_Lt\right)} & 0 \\
0 & -t^*_{Lj}e^{-i\left(\frac{\phi_L}{2}+eV_Lt\right)}           
\end{array} \right)\hspace{1mm}.
\end{equation}
$\hat{G}^r_{ij}(t,t_1)$ and $\hat{G}^<_{ij}(t,t_1)$ are the
retarded and distribution Green's functions in the device normal 
region. They are defined by:
\begin{equation}
\hat{G}^r_{ij}(t,t_1)= -i \theta(t-t_1) 
\left( \begin{array}{ll}
<\{ c_{i\uparrow}(t), c^{\dagger}_{j\uparrow}(t_1)\}> &
<\{ c_{i\uparrow}(t), c_{j\downarrow}(t_1)\}>  \\
<\{ c^{\dagger}_{i\downarrow}(t), c^{\dagger}_{j\uparrow}(t_1)\}> &
<\{ c^{\dagger}_{i\downarrow}(t), c_{j\downarrow}(t_1)\}> 
\end{array}
\right) \hspace{1mm},
\end{equation}
\begin{equation}
\hat{G}^<_{ij}(t,t_1)= i
\left( \begin{array}{ll} 
<c^{\dagger}_{j\uparrow}(t_1) c_{i\uparrow}(t)> &
<c_{j\downarrow}(t_1) c_{i\uparrow}(t)> \\
<c^{\dagger}_{j\uparrow}(t_1) c^{\dagger}_{i\downarrow}(t)> &
<c_{j\downarrow}(t_1) c^{\dagger}_{i\downarrow}(t)> \\
\end{array}
\right) \hspace{1mm}.
\end{equation}
Substituting $\hat{G}^<_{j,Lk}(t,t)$ into Eq.(5), assuming 
$t_{Lj}$ is real, the current $I_L(t)$ can be expressed in terms 
of the Green's functions of the device normal region, as:
\begin{equation}
I_L(t) = -2e Im \int^t_{-\infty} dt_1 \int \frac{d\epsilon}{2\pi}
e^{i\epsilon(t-t_1)} Tr\left\{\left[ 
\tilde{\rho}_L(\epsilon)f_L(\epsilon) \hat{\bf G}^r(t,t_1)
+\beta^*_L(\epsilon) \hat{\bf G}^<(t,t_1)
\right] {\bf \Gamma}_L \hat{\tilde{\Sigma}}_L
\hat{\sigma}_z \right\}\hspace{1mm}, 
\end{equation}
where $f_{L/R}(\epsilon)=\frac{1}{e^{\epsilon/K_BT}+1}$ is 
the Fermi distribution function of electrons in the left/right 
superconducting lead. $\beta_L(\epsilon)$ is defined 
as:\cite{ref21,ref25} 
$\beta_L(\epsilon)=\frac{\epsilon}{i\sqrt{\Delta^2_L
-\epsilon^2}}$ for $\Delta_L>|\epsilon|$, and $\beta_L(\epsilon)
=\frac{|\epsilon|}{\sqrt{\epsilon^2-\Delta^2_L}}$ for
$\Delta_L<|\epsilon|$. $\tilde{\rho}_L(\epsilon)
=Re[\beta_L(\epsilon)] =\theta(|\epsilon|-\Delta)
\frac{|\epsilon|}{\sqrt{\epsilon^2-\Delta^2_L}}$
is the dimensionless BCS density of states, {\it i.e.} the 
ratio of the superconducting density of states 
$\rho^S_L(\epsilon)$ to the normal density of states
$\rho^N_L(\epsilon)$. ${\bf \Gamma}$ is the linewidth matrix function 
defined by $\Gamma_{L;ij}=2\pi t_{Li}t^*_{Lj} \rho^N_L(\epsilon)$,
in which we have assumed that ${\bf \Gamma}_L$ is independent 
energy $\epsilon$.\cite{ref26} In this paper, we use 
boldface letters to denote quantities representing matrices 
whose matrix elements are calculated using states $i,j$ of the
device normal region. Finally, $\hat{\tilde{\Sigma}}_L$ is a compact 
notation,
\begin{equation}
\hat{\tilde{\Sigma}}_L(\epsilon)= \left( \begin{array}{ll}
e^{-ieV_L(t_1-t)} &
-\frac{\Delta}{\epsilon}e^{-i\phi_L-ieV_L(t_1+t)} \\
-\frac{\Delta}{\epsilon}e^{i\phi_L +ieV_L(t_1+t)} &
e^{ieV_L(t_1-t)}
\end{array} \right) \hspace{1mm}.
\end{equation}
The formula Eq.(10) describes the current using Green's functions 
of the normal region. It is a general formula and can therefore
be applied to situations involving arbitrary interactions in 
the normal region and is also applicable at nonequilibrium 
({\it e.g.} at a high bias $V$). If the normal region is
coupled to multiple superconducting leads or to some extra normal 
leads, Eq.(10) is still valid. 

In the following we fix $V_L=0$\cite{ref27} so that the left 
superconducting lead is taken as the potential ground, then 
$\hat{\tilde{\Sigma}}_L$ reduces to:
\begin{equation} 
\hat{\tilde{\Sigma}}_L(\epsilon)= \left( \begin{array}{ll}
1 &
-\frac{\Delta}{\epsilon}e^{-i\phi_L} \\
-\frac{\Delta}{\epsilon}e^{i\phi_L } &
1
\end{array} \right)  \hspace{1mm}.
\end{equation}
Note that the superconducting phase difference between the two 
leads is a time dependent periodic function with a period 
$T=2\pi/\omega$, where $\omega=2eV$ and $V=V_L-V_R$ is the 
bias voltage between the leads. Therefore the time-dependent 
current $I_L(t)$ is also a periodic function with the same 
period $T$ because the Green's functions have the property 
$G(t,t_1)=G(t+T,t_1+T)$.\cite{ref28} Then we can take the 
conventional Fourier expansion for the current $I_L(t)$:
\begin{equation}
I_L(t)=\sum\limits_n I_{Ln}e^{in\omega t} \hspace{1mm},
\end{equation}
and take the double Fourier expansion for the Green's 
function:\cite{ref4,ref26}
\begin{equation}
G(t,t_1)=\sum\limits_n e^{inwt_1} \int \frac{d\epsilon}{2\pi}
e^{-i\epsilon(t-t_1)} G_n(\epsilon)\ \ .
\end{equation}
To simplify notation in the following analysis, we introduce 
quantities $G_{mn}(\epsilon)\equiv G_{n-m}(\epsilon+m\omega)$ and 
${\cal I}_L(t)$,
\begin{equation}
{\cal I}_L(t)= -2e \int^t_{-\infty} dt_1 \int \frac{d\epsilon}{2\pi}
e^{i\epsilon(t-t_1)} Tr\left\{\left[
\tilde{\rho}_L(\epsilon)f_L(\epsilon) \hat{\bf G}^r(t,t_1) 
+\beta^*_L(\epsilon) \hat{\bf G}^<(t,t_1) 
\right] {\bf \Gamma}_L \hat{\tilde{\Sigma}}_L 
\hat{\sigma}_z \right\} \hspace{1mm},
\end{equation}
so that $I_L(t)= Im[{\cal I}_L(t)]$. 

Then the Fourier component of ac current is obtained as:
\begin{equation}
I_{Ln} =\frac{i}{2} ({\cal I}^*_{L,-n}-{\cal I}_{Ln}) \hspace{1mm},
\label{central1}
\end{equation}
and
\begin{equation}
{\cal I}_{Ln} = -2e \int \frac{d\epsilon}{2\pi} Tr \left\{\left[
f_L(\epsilon)\tilde{\rho}_L(\epsilon) \hat{\bf G}_{-n0}^r(\epsilon)
+\frac{1}{2} \beta^*_L(\epsilon) \hat{\bf G}_{-n0}^<(\epsilon)
\right]{\bf \Gamma}_L \hat{\tilde{\Sigma}}_L\hat{\sigma}_z \right\} \ .
\label{central2}
\end{equation}
Eqs. (\ref{central1},\ref{central2}) are the first central results 
of this work. They describe ac components of the time-dependent current 
of a S-normal-S device system in terms of the Fourier
component of the Green's function $\hat{\bf G}_{-n0}^r(\epsilon)$ and 
$\hat{\bf G}_{-n0}^<(\epsilon)$ of the normal region. These formula, 
Eqs.(13), (16), and (17), are valid for arbitrary interactions the
normal region may have, for nonequilibrium situations, and for
devices with other normal leads. They can not, however, be applied
to devices with more than two superconducting leads.

When bias voltage $V$ is zero the current $I_L(t)$
is independent to time $t$, then the current reduces as:  
\begin{equation}
I_{L} = -2e Im \int \frac{d\epsilon}{2\pi} Tr \left\{\left[
f_L(\epsilon)\tilde{\rho}_L(\epsilon) \hat{\bf G}^r(\epsilon) 
+\frac{1}{2} \beta^*_L(\epsilon) \hat{\bf G}^<(\epsilon) 
\right]{\bf \Gamma}_L \hat{\tilde{\Sigma}}_L\hat{\sigma}_z \right\} \ .
\end{equation}

\section{Noninteracting Normal Region}

In this section we apply the general expressions for the ac current
derived above to an example of a S-normal-S device where
the normal region has no electron-electron interactions. For this
situation, the Hamiltonian $H_{cen}$ can be written as:
$$
H_{cen} = \sum\limits_{j,\sigma} \epsilon_{j\sigma} 
c_{j\sigma}^{\dagger}
c_{j\sigma}
+\sum\limits_{i,j,\sigma(i>j)} \left( t_{ij} 
c^{\dagger}_{i\sigma}c_{j\sigma} + H.c. \right)
$$
\begin{equation}
\equiv \sum_{\sigma} H_{cen,\sigma}\ \ .
\label{hcen}
\end{equation}
This Hamiltonian describes a multi-level noninteracting
quantum dot for which $t_{ij}=0$. It also can describe a typical
tight-binding lattice model, in which $t_{ij}\neq 0$, the 
second term in Eq.(\ref{hcen}) denotes the
coupling between different lattice sites.

For the specific $H_{cen}$ of Eq.(\ref{hcen}), we can solve
the Green's functions $\hat{\bf G}^r_{mn}(\epsilon)$ and 
$\hat{\bf G}^<_{mn}(\epsilon)$ using the Dyson equation and the
Keldysh equation: $\hat{\bf G}^r=\hat{\bf g}^r+\hat{\bf G}^r
\hat{\bf \Sigma}^r\hat{\bf g}^r$, and
$\hat{\bf G}^<=\hat{\bf G}^r\hat{\bf \Sigma}^<\hat{\bf G}^a$.
Here $\hat{\bf g}^r$ is the exact Green's function for the device
normal region without coupling to the leads, and it can be easily
derived as:
\begin{equation}
\hat{\bf g}^r(t,t_1)=-i\theta(t-t_1)
\left( \begin{array}{ll}
e^{-iH_{cen\uparrow}(t-t_1)} & 0 \\
0 & e^{i H_{cen\downarrow}(t-t_1)} 
\end{array}
\right)\ \ .
\end{equation}
$\hat{\bf \Sigma}^r$ and $\hat{\bf \Sigma}^<$ are the retarded 
and distribution self-energies due to coupling to the leads, with
$\hat{\bf \Sigma}^{r(<)}(t,t_1)=\hat{\bf \Sigma}^{r(<)}_L(t,t_1) + 
\hat{\bf \Sigma}^{r(<)}_R(t,t_1)$ and
$$ \hat\Sigma^{r}_{L(R),ij}(t,t_1)
=\sum\limits_k \hat{t}^*_{L(R)i}(t) \hat{g}^r_{L(R)k}(t,t_1)
\hat{t}_{L(R)j}(t_1) \hspace{15mm}
$$
\begin{equation}
\mbox{}\hspace{15mm}
=-i\theta(t-t_1)\int\frac{d\epsilon}{2\pi}\Gamma_{L(R),ij}
\beta_{L(R)}(\epsilon)e^{-i\epsilon(t-t_1)}
\hat{\tilde{\Sigma}}_{L(R)} \ \ ,
\end{equation} 
$$ \hat\Sigma^{<}_{L(R),ij}(t,t_1) 
=\sum\limits_k \hat{t}^*_{L(R)i}(t) \hat{g}^<_{L(R)k}(t,t_1) 
\hat{t}_{L(R)j}(t_1) \hspace{15mm} 
$$ 
\begin{equation}
\mbox{}\hspace{15mm}
=i\int\frac{d\epsilon}{2\pi}\Gamma_{L(R),ij}
f_{L(R)}(\epsilon) \tilde{\rho}_{L(R)}(\epsilon)e^{-i\epsilon(t-t_1)}
\hat{\tilde{\Sigma}}_{L(R)} \ \ .
\end{equation}
The Fourier space form of these quantities are easily obtained 
(notice that $V_L=0$ and $V_R=-V$):
\begin{eqnarray}
& & 
\hat{\bf g}^r_{mn}(\epsilon) =\left( \begin{array}{ll}
\delta_{mn}/(\epsilon_m-{\bf H}_{cen\uparrow}+i0^+) & 0 \\
0 & \delta_{mn}/(\epsilon_m +{\bf H}_{cen\uparrow}+i0^+)
\end{array}
\right) \\
& &
\hat{\bf \Sigma}^r_{L;mn}(\epsilon) = -\frac{i}{2} {\bf \Gamma}_L
\delta_{mn} \beta_L(\epsilon_m) 
\hat{\tilde{\Sigma}}_L(\epsilon_m) \\
& &
\hat{\bf \Sigma}^r_{R;mn}(\epsilon) = -\frac{i}{2} {\bf \Gamma}_R
\left(   \begin{array}{ll}
\delta_{mn}\beta_R(\epsilon_{m+\frac{1}{2}}) &
\delta_{m,n-1} \beta_R(\epsilon_{m+\frac{1}{2}}) 
\frac{-\Delta_R}{\epsilon_{m+\frac{1}{2}}} e^{-i\phi_R} \\
\delta_{m,n+1} \beta_R(\epsilon_{m-\frac{1}{2}}) 
\frac{-\Delta_R}{\epsilon_{m-\frac{1}{2}}} e^{i\phi_R}  &
\delta_{mn}\beta_R(\epsilon_{m-\frac{1}{2}})
\end{array} \right)  , \\
& &
\hat{\bf \Sigma}^<_{L;mn}(\epsilon) = i {\bf \Gamma}_L
\delta_{mn} f_L(\epsilon_m)\tilde{\rho}_L(\epsilon_m)
\hat{\tilde{\Sigma}}_L(\epsilon_m)  \\
& &
\hat{\bf \Sigma}^<_{R;mn}(\epsilon) = i {\bf \Gamma}_R
\left( \begin{array}{ll}
\delta_{mn}f_L(\epsilon_{m+\frac{1}{2}})
\tilde{\rho}_R(\epsilon_{m+\frac{1}{2}}) &
\delta_{m,n-1} f_L(\epsilon_{m+\frac{1}{2}})
\tilde{\rho}_R(\epsilon_{m+\frac{1}{2}})
\frac{-\Delta_R}{\epsilon_{m+\frac{1}{2}}}
e^{-i\phi_R}              \\
\delta_{m,n+1} f_L(\epsilon_{m-\frac{1}{2}}) 
\tilde{\rho}_R(\epsilon_{m-\frac{1}{2}}) 
\frac{-\Delta_R}{\epsilon_{m-\frac{1}{2}}} 
e^{i\phi_R}    &
\delta_{mn}f_R(\epsilon_{m-\frac{1}{2}})
\tilde{\rho}_R(\epsilon_{m-\frac{1}{2}}) 
\end{array} \right)  
\end{eqnarray}
where $\epsilon_x=\epsilon+x\omega$.
Similarly, the Fourier space form of the Keldysh equation and the
Dyson equation are:
\begin{eqnarray}
& &
\hat{\bf G}^<_{mn}(\epsilon) = \sum\limits_{l_1,l_2} 
\hat{\bf G}^r_{m l_1}(\epsilon)
\hat{\bf \Sigma}^<_{l_1 l_2}(\epsilon) \hat{\bf G}^a_{l_2n}(\epsilon)
\ \ , \\
& &
\hat{\bf G}^r_{m n}(\epsilon) =\hat{\bf g}^r_{m n}(\epsilon) \delta_{mn} 
+  \sum\limits_l \hat{\bf G}^r_{m l}(\epsilon)
\hat{\bf \Sigma}^r_{l n}(\epsilon)
\hat{\bf g}^r_{n n}(\epsilon) \ \ .
\end{eqnarray}
If $\hat{\bf G}^r_{m n}(\epsilon)$ has been solved, then from the
Keldysh equation (28), $\hat{\bf G}^<_{mn}(\epsilon)$ can be obtained
straightforwardly. Therefore in the following we only need to solve 
the retarded Green's function $\hat{\bf G}^r_{m n}(\epsilon)$. 

From the Dyson equation (29) we have:
\begin{eqnarray}
& &
 {\bf G}_{mn;11}^r ={\bf g}_{mn;11}^r \delta_{mn} +
{\bf G}_{mn;11}^r{\bf \Sigma}_{nn;11}^r{\bf g}_{nn;11}^r 
+\sum\limits_l {\bf G}_{ml;12}^r{\bf \Sigma}_{ln;21}^r{\bf g}_{nn;11}^r
\ \ , \\
& &
 {\bf G}_{mn;12}^r = 
  {\bf G}_{mn;12}^r{\bf \Sigma}_{nn;22}^r{\bf g}_{nn;22}^r
   +\sum\limits_l {\bf G}_{ml;11}^r{\bf \Sigma}_{ln;12}^r{\bf g}_{nn;22}^r
\ \ ,
\end{eqnarray}
where we have suppressed the argument $\epsilon$. From Eq.(31), 
one has:
\begin{equation}
{\bf G}_{mn;12}^r =
\sum\limits_l {\bf G}_{ml;11}^r{\bf \Sigma}_{ln;12}^r 
\frac{1}{{\bf g}^{r-1}_{nn;22}- {\bf \Sigma}_{nn;22}^r}\ \ .
\end{equation}
Substituting this expression to Eq.(30) one easily finds:
\begin{equation} 
 {\bf G}_{mn;11}^r = \frac{\delta_{mn}}{{\bf g}^{r-1}_{nn;22}- 
{\bf \Sigma}_{nn;22}^r }
+ \sum\limits_l {\bf G}_{ml;11}^r  {\bf B}_{ln}
\ \ ,
\end{equation}
where
\begin{equation}
 {\bf B}_{mn}(\epsilon) \equiv \sum\limits_l 
{\bf \Sigma}^r_{ml;12} 
\frac{1}{{\bf g}^{r-1}_{ll;22}- {\bf \Sigma}_{ll;22}^r}
{\bf \Sigma}^r_{ln;21}
\frac{1}{{\bf g}^{r-1}_{nn;11}- {\bf \Sigma}_{nn;11}^r}\ \ .
\end{equation} 
Note ${\bf B}_{mn}\neq 0$ only when $m=n,n\pm1$.
The quantity ${\bf B}_{mn}$ has a clear physical meaning: it
describes the intensity of Andreev reflection processes, an 
example is shown in Fig.1 in which a particle in the normal 
region undergoes twice Andreev reflections. Then by iterating 
Eq.(33), ${\bf G}_{mn;11}^r$ can be formally solved,   
\begin{equation}
 {\bf G}_{mn;11}^r = \frac{\delta_{mn}}{{\bf g}^{r-1}_{nn;11}-
{\bf \Sigma}_{nn;11}^r }
+ \frac{1}{{\bf g}^{r-1}_{mm;11}-{\bf \Sigma}_{mm;11}^r }
{\bf Y}_{mn} \ \ ,
\end{equation}
where
$$
{\bf Y}_{mn}= {\bf B}_{mn} + \sum\limits_{l_1}{\bf B}_{ml_1} {\bf B}_{l_1n}
+ \sum\limits_{l_1, l_2}{\bf B}_{ml_1} {\bf B}_{l_1 l_2}{\bf B}_{l_2n} 
+ ...
$$
\begin{equation}
= {\bf B}_{mn} +\sum\limits_l {\bf B}_{ml} {\bf Y}_{ln} \ \ .
\end{equation}
Similarly, the quantity ${\bf Y}_{mn}(\epsilon)$ has a clear physical
meaning: it gives the intensity of the process for which
an electron having initial energy $\epsilon+n\omega$ ends up with
final energy $\epsilon+m\omega$ after going through all possible multiple 
Andreev reflections in the normal region. Eq.(36) can only be solved 
numerically and after ${\bf Y}_{mn}$ is solved, from Eqs.(35) and (32) 
${\bf G}_{mn;11}^r$ and ${\bf G}_{mn;12}^r$ can be obtained immediately.
Finally, ${\bf G}_{mn;21}^r$ and ${\bf G}_{mn;22}^r$ can also be
calculated using following equations which are derived from the 
Dyson equation:
\begin{eqnarray}
& &
{\bf G}_{mn;21}^r = \sum\limits_l 
\frac{1}{{\bf g}_{mm;22}^{r-1}- {\bf \Sigma}_{mm;22}^r }
{\bf \Sigma}^r_{ml;21} {\bf G}^r_{ln;11} \ \ , \\
& &
{\bf G}_{mn;22}^r = 
\frac{\delta_{mn}}{{\bf g}_{mm;22}^{r-1}- {\bf \Sigma}_{mm;22}^r }
+ \sum\limits_l 
\frac{1}{{\bf g}_{mm;22}^{r-1}- {\bf \Sigma}_{mm;22}^r }
{\bf \Sigma}^r_{ml;21} {\bf G}^r_{ln;12}\ \ .
\end{eqnarray}
With $\hat{\bf G}_{mn}^r$ and $\hat{\bf G}_{mn}^<$ solved, from Eq.(17) 
the ac component and time-dependent current can be calculated 
without further complications.

In the rest of this section, we present numerical results for which
some further simplifications are made. We reduce the device normal region
to a quantum dot with a spin degenerate single level, 
{\it i.e.} $H_{cen} =\sum\limits_{\sigma} \epsilon_d 
c^{\dagger}_{\sigma}
c_{\sigma}$. For this case the boldface matrices reduce to a C number. 
We also take $\Delta=\Delta_L=\Delta_R=1$ as the energy unit and only 
consider devices with symmetric barriers ($\Gamma_L=\Gamma_R$). 
It should be mentioned that since we have assumed a spin independent 
intradot level $\epsilon_d$ and hopping elements $t_{L(R)}$, 
$<c^{\dagger}_{\uparrow}(t_1) c_{\uparrow}(t)>$ should be equal
to $<c^{\dagger}_{\downarrow}(t_1) c_{\downarrow}(t)>$. Following this
we have $G^<_{11}(t,t_1) +G^<_{22}(t_1,t) = -\left[
G^r_{11}(t,t_1) - G^a_{11}(t,t_1)\right]$ and $\hat{G}^<(t,t_1)
=-[\hat{G}^<(t_1,t)]^{\dagger}$.\cite{ref29} The Fourier forms are 
$G^<_{mn;11}(\epsilon)+G^<_{-n,-m;22}(-\epsilon) = - [ 
 G^r_{mn;11}(\epsilon) -G^{r*}_{nm;11}(\epsilon) ]$
and $\hat{G}_{nm}^<(\epsilon) =-[\hat{G}^<_{mn}(\epsilon)]^{\dagger}$.
These relationships provide very strong checks on our analytical
derivations and numerical calculations which we present in the following
subsections.

\subsection{Intradot distribution of electrons}

In this subsection we present results of the intradot 
distribution of electrons for the S-normal-S device.
Because of the finite bias voltage $V$, the current, intradot
occupation number of electrons, local density of states (LDOS), 
and the intradot distribution of electrons, are all functions
of time $t$. The time average occupation number of electrons 
on the intradot state $\uparrow$ is: (same for state $\downarrow$)
\begin{equation}
<n_{\uparrow}(t)>_t =-i <G_{11}^<(t,t)>_t = -i \int \frac{d\epsilon}{2\pi}
G_{00;11}^<(\epsilon)\ \ .
\label{nup}
\end{equation}
The integrand of (\ref{nup}), $\frac{-i}{2\pi} G_{00;11}^<(\epsilon)$, 
is the time-averaged occupation number of electrons with energy $\epsilon$.
Here, subscript ``11" are indexes of the $2\times2$ Nambu matrix element, 
and ``00" are indexes of Fourier component.
The average LDOS is given by $LDOS(\epsilon)=-\frac{1}{\pi} Im[
G^r_{00;11}(\epsilon)+G^r_{00;22}(-\epsilon)]$. The average intradot
distribution of electrons can be obtained from the average occupation number at
energy $\epsilon$ and the average $LDOS(\epsilon)$\cite{ref30},
\begin{equation}
f_d(\epsilon) = 
\frac{iG^<_{00;11}(\epsilon)}{2Im\left[G^r_{00;11}(\epsilon)\right]}\ .
\label{fd}
\end{equation}
It is important to emphasize that the distribution of electrons can be
experimentally measured\cite{ref31,ref32}. For example, recently Pierre 
{\sl et. al.} have measured\cite{ref32} this distribution for a S-normal-S 
device where the normal region is a diffusive mesoscopic metallic wire. 
They reported a multiple step structure for the distribution of
electrons 
in that device\cite{ref32}.

Fig.2 shows the average intradot distribution of electrons at different bias
voltage $V$ for our system with a very large coupling $\Gamma$. When $\Gamma$ 
is large, coupling between the superconducting leads and the normal region is
strong, therefore the device behaves like a S-ballistic-normal-conductor-S 
system.  The property of the electron distribution in this situation is the 
following. When $min(V_L-\Delta, V_R-\Delta)<\epsilon<max(V_L+\Delta, 
V_R+\Delta)$, the distribution is a constant, {\it i.e.}
$f_d(\epsilon)\sim 1/2$ for symmetric couplings. When $\epsilon$ goes away 
from this region, the distribution quickly rises (or drops) to unity (or to 
zero) for $\epsilon<min(V_L-\Delta,V_R-\Delta)$ 
(or for $\epsilon>max(V_L+\Delta,V_R+\Delta$ )). 

To contrast with the experimental results of Pierre {\sl et. al.}\cite{ref32}, 
here the distribution is a constant instead of the multiple step structure 
between the gap, even though multiple Andreev reflections do occur in our
system. This difference originates from the different property of the
central device region, {\it i.e.}, our normal region is ballistic while that 
in Pierre {\sl et. al.} experiment is diffusive\cite{ref32}. In order to 
explain it in more detail, the inset of Fig.2 shows a particular multiple
(two) Andreev reflection process. To start, an incident electron at
$A_i$ below the gap of the left lead tunnels into the QD, it passes two Andreev 
reflections (through the points labelled as A1-A6) inside the QD and finally 
tunnels into the right lead (at $A_e$) which is higher than the 
gap of the right lead. Due to the ballistic nature of the QD, the distribution 
of electrons at point A1 is the same as at A2, the distribution of holes 
at A3 is the same as at A4, while distribution of electrons at A5 is the same 
as at A6. When $\Gamma$ is large, the probability of Andreev reflection inside 
the QD within the energy gap is very close to unity\cite{ref33}, and hence the 
hole distribution at A3 is, to a very good extent, the same as the distribution
of electrons at A2. Similarly the hole distribution at A4 is approximately
the same as the electron distribution at A5. We hence conclude that 
for the ballistic normal region, the distribution of particles (electrons and 
holes) along this path is the same everywhere, except at the abrupt change 
during the tunneling process at $A_i$ and $A_e$ from and to the two leads.  
Moreover, for symmetric barriers, the distribution of particles along the 
A1-A6 path must be $1/2$. This explains why we obtained a constant $1/2$ 
distribution at $min(V_L-\Delta,V_R-\Delta)<\epsilon<max(V_L+\Delta,
V_R+\Delta)$ as shown in Fig.2. This also explains why we expect a different 
distribution when the normal region is diffusive: for a diffusive conductor 
the distribution at A1 and A2 must be different due to diffusive scattering 
between the two points, therefore the distribution of particles will 
continuously vary from one to zero along the path A1-A6.

Next, we investigate the distribution of electrons for $\Gamma\sim\Delta$,
the results are shown in Fig.3. For this case, a most prominent behavior 
of $f_d(\epsilon)$ is that it oscillates as a function of $\epsilon$. 
The oscillations also become more rapid when bias voltage $V$ is reduced.
An oscillatory $f_d(\epsilon)$ means its value is not necessarily smaller 
for larger $\epsilon$, hence a ``population inversion'' is possible. 
This population inversion originates from the non-monotonic probability 
of Andreev reflections. For example, $f_d(\epsilon)$ has a dip at 
$\epsilon=V_R-\Delta$, due to the following reason. For an incident electron 
coming from the left lead with energy $V_R-\Delta$, this electron has a small 
but nonzero probability to pass the left barrier. After tunneling through, 
it reaches the right barrier where an Andreev reflection occurs.
Because this electron has energy $\epsilon=V_R-\Delta$, the Andreev 
reflection occurs with probability one\cite{ref33}. 
Therefore the distribution of electrons at this energy $\epsilon$ is very 
small. When $\epsilon$ deviates from $V_R-\Delta$, the probability of 
Andreev reflection decreases leading to a larger $f_d$, hence we
expect a dip in $f_d$ to emerge at $\epsilon=V_R-\Delta$.

\subsection{Local density of states}

In this subsection, we investigate another important quantity, the
LDOS. We will mainly discusses Andreev bound states at a finite bias $V$. 
If bias $V>2\Delta$, multiple Andreev reflections are very weak hence no
Andreev bound states can form in the QD. In this case the intradot level 
$\epsilon_d$ is only slightly shifted due to a non-zero real part of 
the self-energy $\Sigma^r$, the level half-width is still on the scale
of $\Gamma_{L/R}$, and an extra structures (a dip and a peak) emerge
in the curves of $LDOS(\epsilon)$ versus $\epsilon$ at the superconducting 
gap (not shown in here). 

Much more interesting is the case of $V<\Delta$, shown in Fig.4 at different 
bias $V$. A series of very narrow peaks emerge in $LDOS(\epsilon)$, clearly
indicate the formation of Andreev bound states inside the QD.
Note that they are not rigorous bound states but are quasi-bound states with
a finite life time, and after many Andreev reflections the particle can leave 
the QD. This is different from the zero bias situation\cite{ref17}. The 
half-width of Andreev bound states is much narrower than $\Gamma$. With 
a decreasing bias $V$, they become even narrower with a higher intensity.
The average distance between two successive Andreev bound states is about 
$eV$. When $neV$ and $(n+1)eV$ ($n=0,\pm1,\pm2$,...) are within the gap,
there exists an Andreev bound state between $\epsilon=neV$ and $(n+1)eV$. 
Moreover, these Andreev bound states are symmetrically distributed 
at the two side of $V_L$ and $V_R$. This means the following: when an 
incident electron from below the gap aligns perfectly with an Andreev bound 
state of the QD, even after many Andreev reflections it will always stay 
on the Andreev bound state until it leaves the QD (see inset of Fig.4(a)). 
Along this path, the particle goes through all Andreev bound states, and 
a resonance multiple Andreev reflection occurs. Occasionally, a quenching 
of Andreev bound state is observed to occur. In this case, a specific 
Andreev bound state may have very low LDOS at a specific bias $V$, an 
example is indicated by the arrow in Fig.4(b). 

The results of Fig.4 is obtained by fixing the intradot level $\epsilon_d$ to 
zero ({\it i.e.} at the center of the gap).  Next, we investigate 
how are Andreev bound states affected when $\epsilon_d\neq 0$, the results
shown in Fig.5. With $\epsilon_d\neq 0$, the Andreev bound states are shifted
in their positions, but their physical characteristics are the same as
those of $\epsilon_d=0$. The amount of shift is not $\epsilon_d$ but much 
smaller and two successive Andreev bound states are shifted in opposite 
directions. If an Andreev bound state is in the energy range from 
$\epsilon=neV$ to $(n+1)eV$, it stays in this range at any value of 
$\epsilon_d$. Their heights vary with $\epsilon_d$, when $\epsilon_d$ is
in the range of $neV$ to $(n+1)eV$, the peak in this range reaches a maximum
value.

An important property of the Andreev bound states is their ability to carry 
current. From Eqs.(16) and (17), the time-averaged current density 
$j_0(\epsilon)$ is obtained to be:
\begin{equation}
j_0(\epsilon)=-\frac{e}{\pi} Im Tr \left\{\left[ 
f_L(\epsilon)\tilde{\rho}_L(\epsilon) \hat{G}_{00}^r(\epsilon) +
\frac{1}{2} \beta_L^*(\epsilon) \hat{G}^<_{00}(\epsilon) \right]
\Gamma_L \hat{\tilde{\Sigma}}_L \hat{\sigma}_z \right\}\ .
\end{equation}
The current density is related to time-averaged current as
$I_{0}=\int d\epsilon j_0(\epsilon)$. 
In Fig.6, we show intradot distribution of electrons 
$f_d$ (solid curve in Fig.6a), LDOS (dotted curve in Fig.6a), and the 
time-averaged current density (Fig.6b) $j_0(\epsilon)$. Several observations 
are in order. (i). Although $f_d(\epsilon)$ is oscillating between $0$ and 
$1$ in a complicated manner, its value at each Andreev bound state (the peak
positions of the dotted curve) is alway $1/2$. This is because resonant
multiple Andreev reflections occur along the path of Andreev bound states (as
shown in the inset of Fig.4(a)). (ii). The current density $j_0(\epsilon)$ is
dominated by a series of peaks located precisely at the energies of 
Andreev bound states. This is a clear indication that current is carried by
Andreev bound states. When $min(V_L-\Delta,V_R-\Delta)<
\epsilon<max(V_L+\Delta,V_R+\Delta)$, the peaks of $j_0(\epsilon)$ all have 
the same height: this means each Andreev bound state carries exactly the 
same amount of current in the same flow direction. The reason for this peculiar
behavior is simple. Along the path of Andreev bound states (inset of Fig.4(a)),
all the electrons move in one direction while all the holes move in opposite 
direction, and along any one path the particle current must be same
everywhere.
Therefore the Andreev bound states carry same amount of current in the 
same direction. This property is qualitatively different from that of the
zero bias case\cite{ref1,ref17}, in which the successive Andreev bound states
carry current with opposite sign.

\subsection{The current}

The time-averaged current $I_0$ of S-normal-S systems has been extensively 
investigated both theoretically and experimentally. A main characteristic of 
the I-V curve $I_0(V)$ is its subharmonic gap structure at 
$V=2\Delta/n$\cite{ref3,ref4,ref5,ref6,ref7,ref8,ref34}, our results are 
shown in Fig.7. The I-V curves also exhibit subharmonic gap structure with a
concomitant 
appearance of negative differential conductance. These results are in 
agreement with those reported recently by Yeyati {\sl et. al.}\cite{ref6} 
and Johansson {\sl et. al.}\cite{ref10}. In the following, we focus on the 
ac component of the current.

From Eqs.(13) and (16), we decompose the time-dependent current into its 
dissipative contribution $I_n^c$, and nondissipative contribution 
$I_n^s$\cite{ref4},
\begin{equation}
I_L(t)= I_{0} +\sum\limits_n I^c_{L0} \cos n\omega t +
\sum\limits_n I^s_{L0} \sin n\omega t \hspace{1mm},
\end{equation}
where $I_{Ln}^c \equiv Im ({\cal I}_{Ln} +{\cal I}_{L-n})$ and
$I_{Ln}^s \equiv Re ({\cal I}_{Ln} -{\cal I}_{L-n})$. Fig.8 and Fig.9 show
the first and second ac components of $I_{Ln}^c$ and $I_{Ln}^s$ as a function 
of bias $V$, and they are marked by a strong oscillatory behavior. The period
of oscillations is roughly given by $\frac{V^2}{\Delta}$, which is dependent
on bias $V$. Generally,
for $\frac{2\Delta}{m}< eV < \frac{2\Delta}{m+1}$ (m=1,2,...), we found that
the ac components oscillate from a maximum to a minimum or vise versa.
When $V>\frac{2\Delta}{n}$, the components $I_{Ln}^c$ and $I_{Ln}^s$ 
quickly decay to zero. When $eV\sim \frac{\Delta}{n}$, the amplitudes of 
the oscillations reach maximum. At $eV\rightarrow 0$, $I_{Ln}^c$ decays
to zero while $I_{Ln}^s$ keeps a finite value. These behaviors are different
from those devices whose normal region has no electronic structure. For 
instance, the result of S-QPC-S system shows no oscillation\cite{ref4}.
 
The time-dependent current $I_L(t)$ is shown in Fig.10. $I_L(t)$ is a well 
known oscillatory function of time $t$ with a frequency $\omega=2eV$. 
When bias $V$ is large, $eV>\Delta$, the high-order Fourier components 
have negligible contribution and $I_L(t)$ can be approximated by
$I_L(t)\approx I_{0}+ I_{L1} sin(\omega t +\phi)$. On the other hand, when
$V$ is small, high-order components contribution substantially and
$I_L(t)$ deviates from a simple sine-like curve.

\section{Conclusions}

In this work, we have derived a general formula for ac components 
of the time-dependent current of arbitrary ballistic S-normal-S systems
where the normal region has its own electronic structure.  The formula 
(Eq.(17)) is valid even when there is a strong interaction in the normal
region of the hybrid device. We then applied this result to study 
ac Josephson current for a system with the normal region being a noninteracting 
single level quantum dot. The average intradot distribution of electrons, 
the average intradot density of states, and ac components of the 
time-dependent current are investigated in detail. The distribution 
exhibits an interesting population inversion, a result that is qualitatively 
different from that of the diffusive normal region. A series of Andreev 
bound states are formed at bias $V<\Delta$ in our system. The peak heights
of LDOS for these Andreev bound states are not the same, but each
state carries the same amount of current. The distribution of electrons at 
the Andreev bound states are all the same, {\it e.g.} equals 1/2 for 
symmetric tunnel barriers. In general, the ac components of the time-dependent 
current has an oscillatory behavior against bias. Depending on the value of 
bias, the high-order ac components may or may not contribute to the total
time-dependent current, leading to a non-sine-like or a sine-like 
dependence on time for the total current.

Finally, we comment on the $eV\rightarrow 0$ limit for the S-QD-S system of 
this work. While our general current formula, Eq.(17), is valid for any bias, 
how to correctly include important physical factors in an actual computation 
of the various quantities of Eq.(17), needs to be discussed. When bias is very 
small, $eV\ll \Delta$, an incident electron from below the gap of the left 
superconducting lead undergoes many Andreev reflections in the QD so as to
go above the gap of the right superconducting lead before exiting the QD.
Therefore the dwell time $\tau_p$ of the particle in the QD becomes long. At 
the limit $eV\rightarrow 0$, $\tau_p$ tends to large values. When $\tau_p$ is 
larger than the mean inelastic scattering time, the intradot relaxation 
effect should be considered in calculating the Green's functions involved in
Eq.(17). When there is no electronic structure in the normal region of the 
device, for instance in a S-QPC-S system\cite{ref4,ref5}, the 
$eV\rightarrow 0$ limit has a variety of different regimes depending on an 
inelastic scattering rate parameter $\delta$ and a transmission probability 
of the QPC\cite{ref4,ref5}. For our S-QD-S system, while relaxation in 
the leads can similarly be included by introducing the same parameter
$\delta$ into the Green's function of the leads\cite{ref4}, this simple 
phenomenological approach can not be applied in the normal QD region.
This is because distribution of leads is determined by their chemical 
potential, however the distribution in the QD must be calculated 
self-consistently for our system. Indeed, if one introduces a finite $\delta$ 
in the QD Green's function, current conservation will be violated. 
A proper treatment of this problem is, perhaps, to explicitly introduce an 
electron-phonon interaction term in the Hamiltonian. This is a very 
complicated problem to solve and we hope to be able to report such an 
analysis in the future.

\section*{ACKNOWLEDGMENTS}

We gratefully acknowledge financial support from NSERC of Canada, FCAR of
Quebec (Q.S. and H.G.), and for a RGC grant from the SAR Government of
Hong Kong under grant number HKU 7215/99P (J.W.). Q.S. thanks Y. Liu and 
X.B. Zhu for their help on the numerical calculations.

\newpage

\section*{Figure Captions}

\begin{itemize}
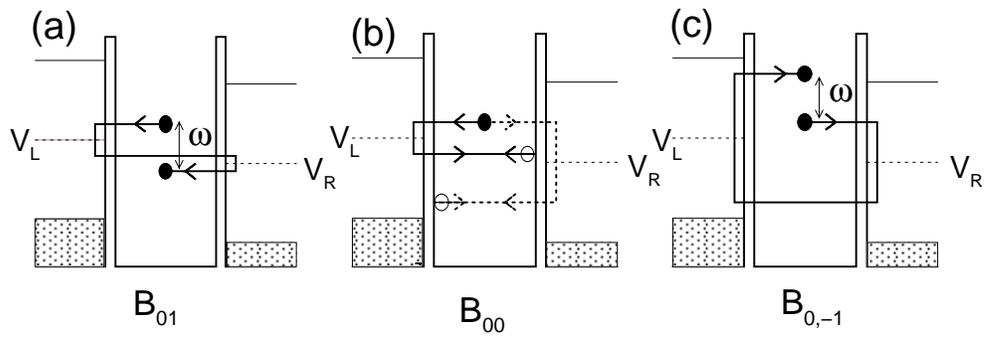
\item[{\bf Fig. 1}]   
A schematic diagram for the transport process consisting of two Andreev
reflections. (a). The particle is first Andreev reflected by the
left superconducting lead, then it is by the right superconducting lead.
This is described by the quantity ${\bf B}_{01}(\epsilon)$. After this 
process, the particle energy reduces by $2eV$ ({\it i.e.} $\omega=2eV$).
(b). The particle is first Andreev reflected by the left (right) lead, followed
by another reflection at the same lead. This process is described by 
quantity ${\bf B}_{00}(\epsilon)$. After this process, the particle energy 
does not change. (c). The particle is first Andreev reflected by
the right lead, then by the left lead. It is described by 
quantity ${\bf B}_{0,-1}(\epsilon)$. After this process, the particle
energy rises $2eV$. All processes with an even number Andreev reflections 
can be decomposed to the three processes plotted here. All processes with
an odd number of Andreev reflections can be decomposed to the even case plus 
one more reflection.

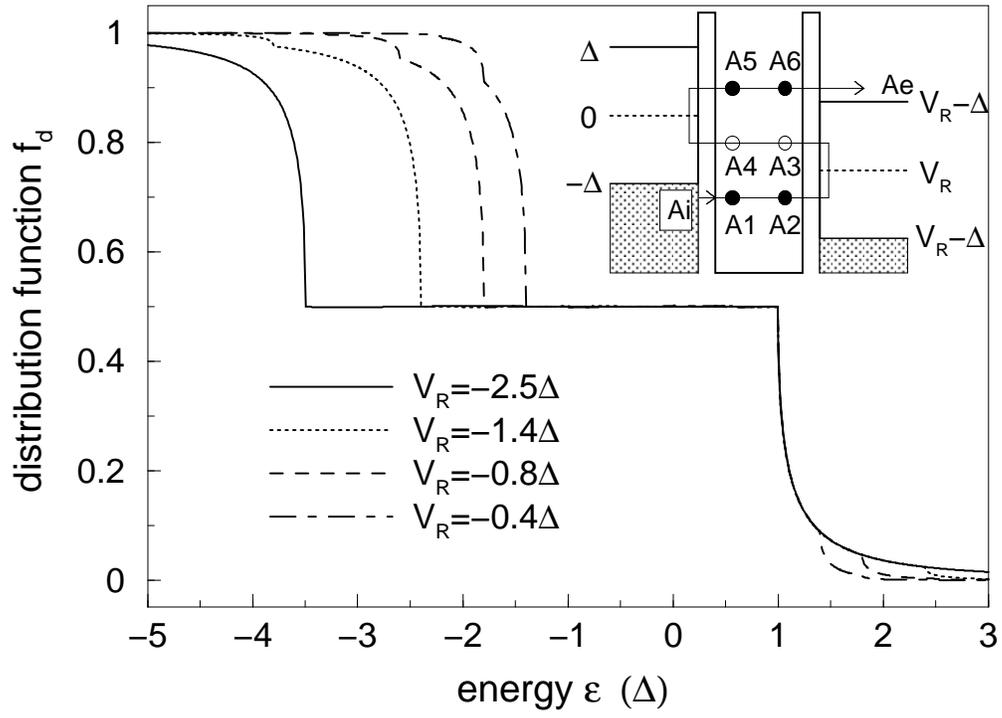
\item[{\bf Fig. 2}]
The time-averaged intradot distribution of electrons versus energy 
$\epsilon$ at large $\Gamma$, $\Gamma_L=\Gamma_R=1000\Delta$. Temperature 
$K_BT=0.05\Delta$, $\epsilon_d=0$, $\delta=0$
($\delta$ is the inelastic scattering rate in two superconducting leads),
and $\phi_L=\phi_R=0$.
Note the fact that the time-averaged distribution, LDOS, and the ac 
components of the current are all independent with initial values of
$\phi_L$ and $\phi_R$ at $\delta=0$. Inset: schematic diagram showing
a multiple (two) Andreev reflection process.

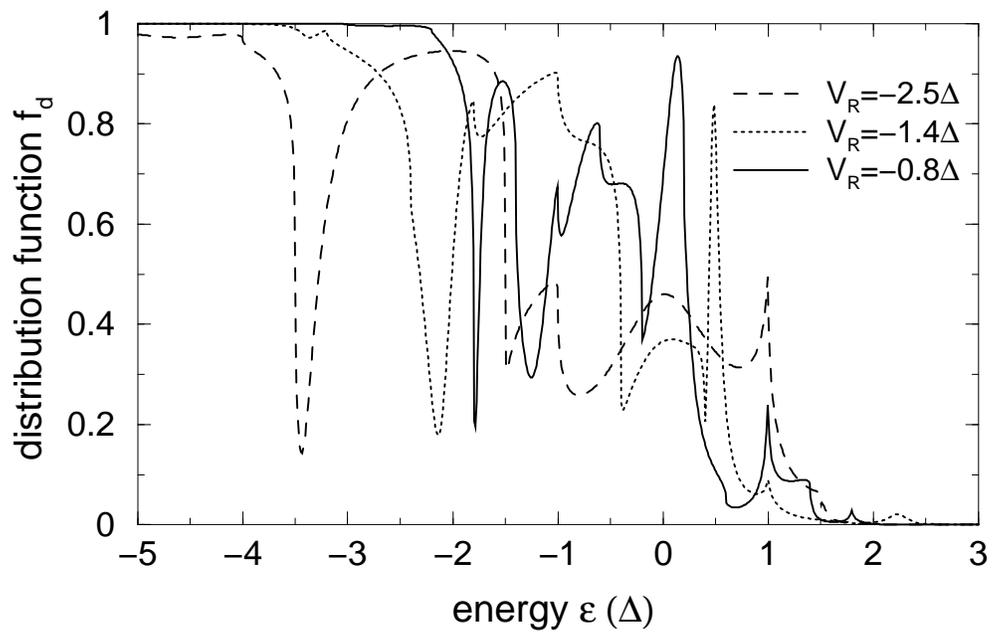
\item[{\bf Fig. 3}]
The time-averaged intradot distribution of electrons versus energy 
$\epsilon$ at general $QD$ parameters, $\Gamma_L=\Gamma_R=1.5\Delta$. 
Other parameters are the same as those of Fig.2.

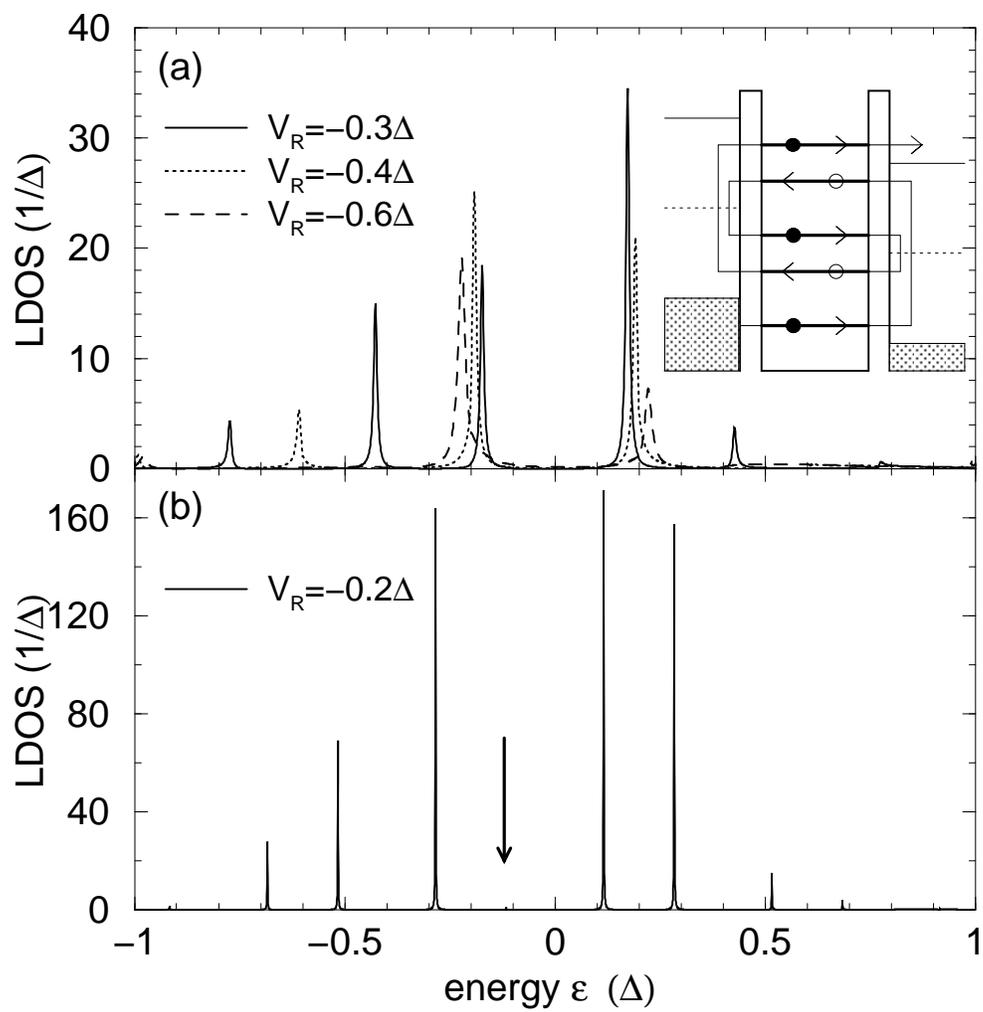
\item[{\bf Fig. 4}]
The time-averaged LDOS versus $\epsilon$ at different bias $V$. 
$K_BT=0.1\Delta$, $\Gamma_L=\Gamma_R=0.8\Delta$, $\epsilon_d=0$, 
and $\delta=0$.  The downward arrow in (b) points to an Andreev
bound state with a very small LDOS. Inset in (a): schematic diagram showing 
a multiple Andreev reflection which passes through the Andreev bound 
states indicated by the thick solid lines in the QD. 

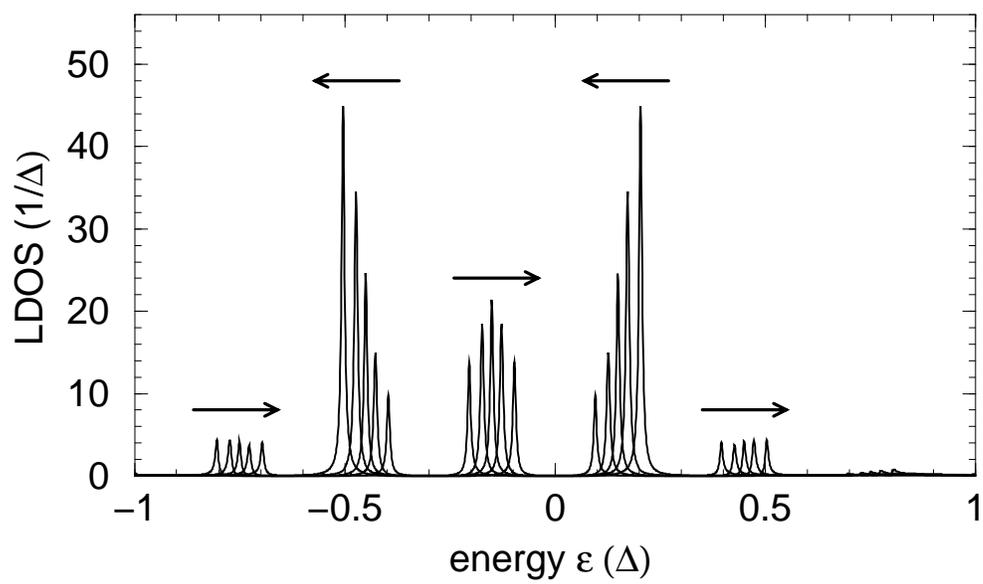
\item[{\bf Fig. 5}]
The time-averaged LDOS versus $\epsilon$ at different level positions
$\epsilon_d$. $V_R=-0.3\Delta$ and other parameters are same as those of
Fig.4. Different curves correspond to $\epsilon_d=0.15\Delta$, $0\Delta$, 
$-0.15\Delta$, $-0.30\Delta$, and $-0.45\Delta$, along the arrow direction.

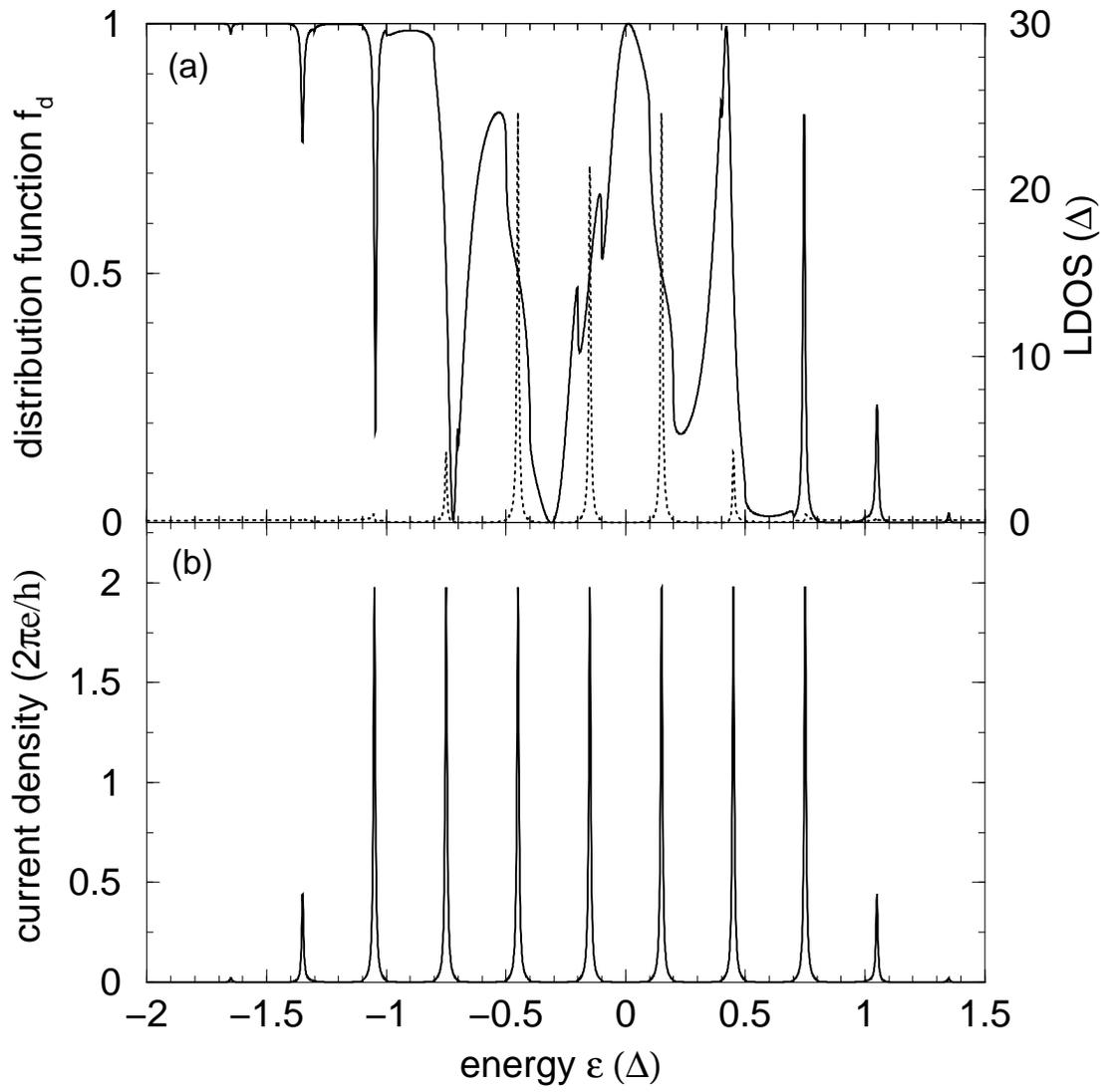
\item[{\bf Fig. 6}]
(a). The time-averaged LDOS (dotted) and the time-averaged distribution 
of electrons (solid) versus $\epsilon$; (b) the time-averaged current
density. $\epsilon_d=-0.15$ and other parameters are same as those in
Fig.5.

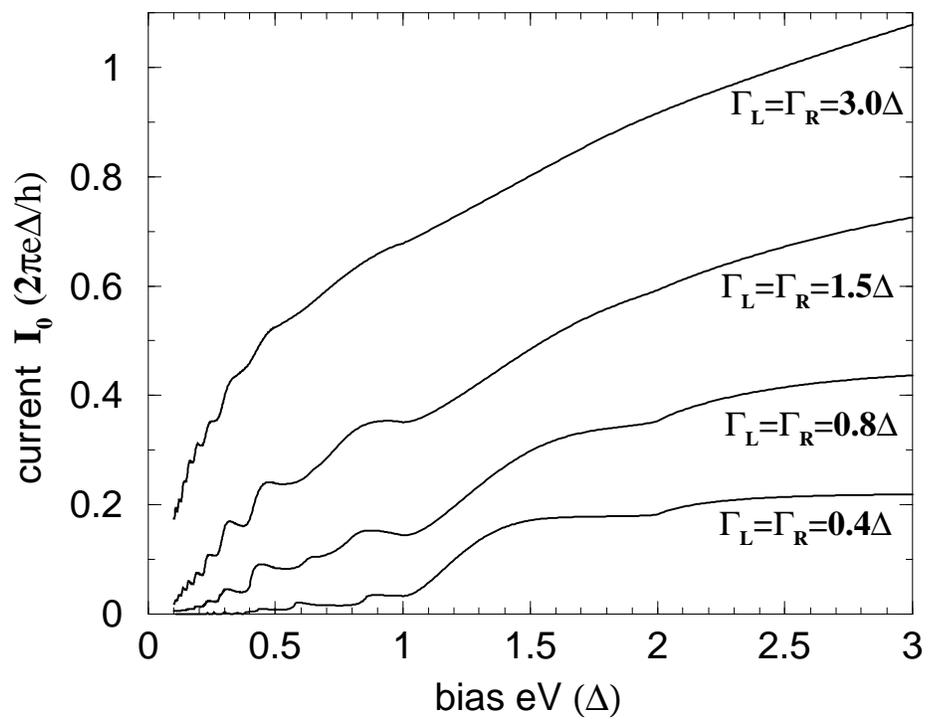
\item[{\bf Fig. 7}]
The time-averaged current $I_0$ versus bias $V$ at different $\Gamma$.
Other parameters:
$K_BT=0.1\Delta$, $\epsilon_d=0$, $\delta=0.005\Delta$, $\phi_L=\phi_R=0$.

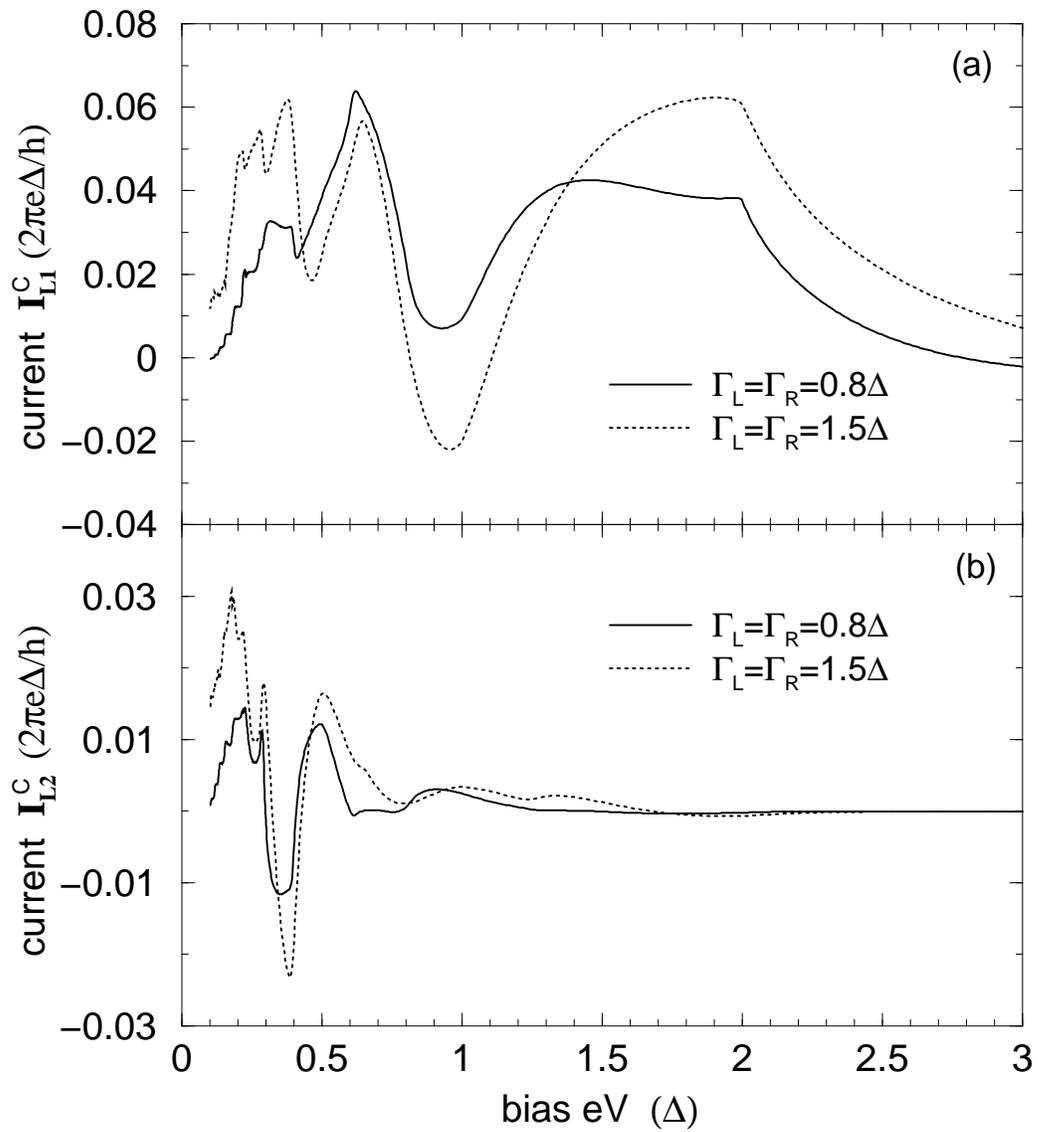
\item[{\bf Fig. 8}]
The dissipative ac components $I_{L1}^c$ and $I_{L2}^c$ versus bias $V$ at
different $\Gamma$. Other parameters are the same as those of Fig.7.

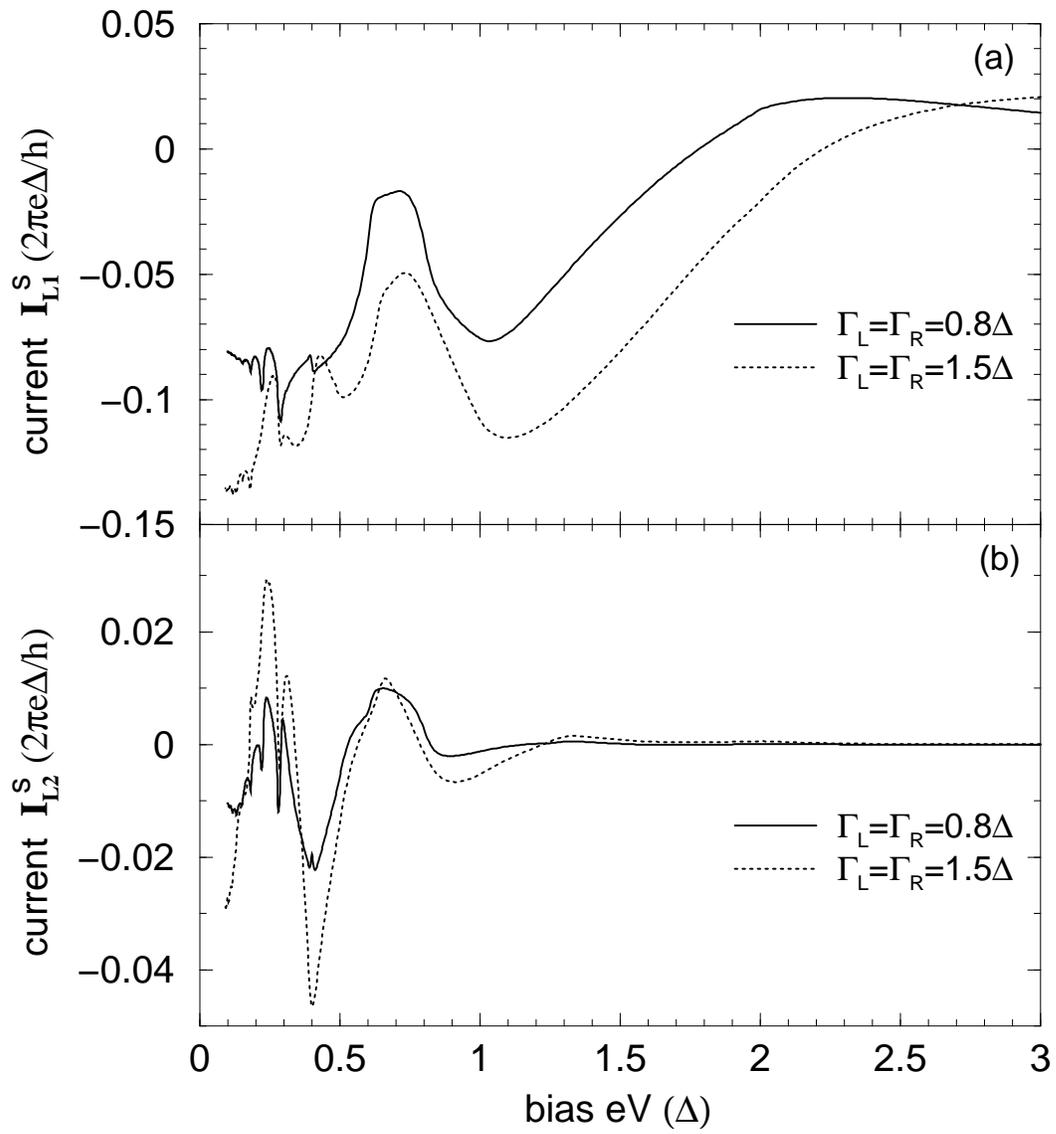
\item[{\bf Fig. 9}]
The nondissipative ac components $I_{L1}^s$ and $I_{L2}^s$ versus bias $V$ at
different $\Gamma$. Other parameters are the same as those of Fig.7.

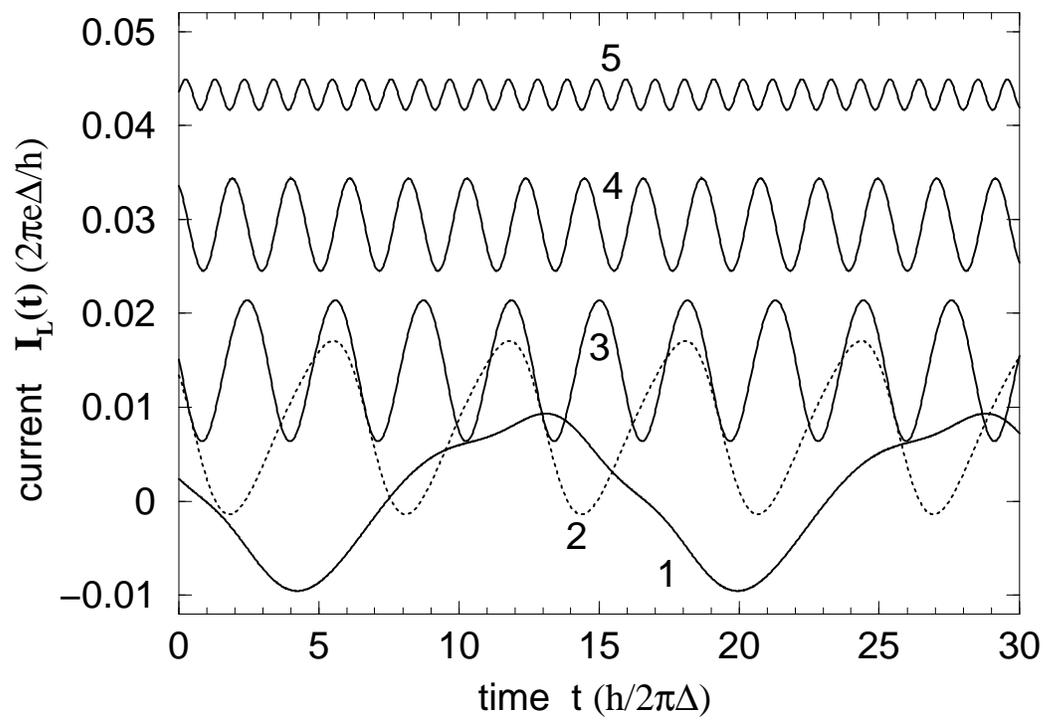
\item[{\bf Fig. 10}]
Time-dependent current $I_L(t)$ versus time $t$ at different bias $V$.
$\Gamma_L=\Gamma_R=0.8\Delta$ and other parameters are the same as those
of Fig.7. The curves labelled $1$ to $5$ correspond to $V=-V_R=0.2\Delta$, 
$0.5\Delta$, $1.0\Delta$, $1.5\Delta$, and $3.0\Delta$, respectively.


\end{itemize}


\begin{references}

\bibitem{ref1}
B.J. van Wees and H. Takayanagi, in {\sl Mesoscopic Electron Transport}, 
edited by L.L. Sohn, L.P. Kouwenhoven, and G. Sch$\ddot{o}$n 
(Kluwer, Dordrecht, 1997).

\bibitem{ref2}
Mesoscopic Superconductivity, edited by F.W.J. Hekking, G. Sch$\ddot{o}$n, and
D.V. Averin [Physica B {\bf 203}, 201 (1994)].

\bibitem{ref3}
E.N. Bratus, V.S. Shumeiko, and G. Wendin, Phys. Rev. Lett. {\bf 74}, 
2110 (1995).

\bibitem{ref4}
J.C. Cuevas, A. Martin-Rodero, and A.L. Yeyati, Phys. Rev. B {\bf 54}, 7366
(1996).

\bibitem{ref5}
E.N. Bratus, V.S. Shumeiko, E.V. Bezuglyi, and G. Wendin, Phys. Rev. B 
{\bf 55}, 12666 (1997).

\bibitem{ref6}
A.L. Yeyati, J.C. Cuevas, A. Lopez-Davalos, and A. Martin-Rodero, Phys. Rev. B
{\bf 55}, R6137 (1997).

\bibitem{ref7}
T.I. Ivanov, Phys. Rev. B {\bf 59}, 169 (1999).

\bibitem{ref8}
K. Kang, Phys. Rev. B {\bf 57}, 11891 (1998).

\bibitem{ref9}
S. Ishizaka, J. Sone, T. Ando, Phys. Rev. B {\bf 52}, 8358 (1995).

\bibitem{ref10}
G. Johansson, E.N. Bratus, V.S. Shumeiko, and G. Wendin, Phys. Rev. B {\bf 60},
1382 (1999).

\bibitem{ref11}
A. Golub and B. Horovitz, Phys. Rev. B {\bf 50}, 15882 (1994).
 
\bibitem{ref14}
A. Yu. Kasumov, {\sl et. al.}, Science {\bf 284}, 1508 (1999).

\bibitem{ref15}
A.F. Morpurgo, J. Kong, C.M. Marcus, and H. Dai, Science {\bf 286}, 263 (1999).

\bibitem{ref16}
Y. Wei, J. Wang, H. Guo, H. Mehrez, and C. Roland, Phys. Rev. B {\bf 63}, 
195412 (2001).

\bibitem{ref17}
P.F. Bagwell, Phys. Rev. B {\bf 46}, 12573 (1992).

\bibitem{ref18}
J.J.A. Baselmans, A.F. Morpurgo, B.J. van Wees, and 
T.M. Klapwijk, Nature (London) {\bf 397}, 43 (1999).

\bibitem{ref19}
E. Scheer, {\sl et. al.}, Nature (London) {\bf 394}, 154 (1998).

\bibitem{ref20}
E. Scheer, W. Belzig, Y. Naveh, M.H. Devoret, D. Esteve, and C. Urbina, 
Phys. Rev. Lett. {\bf 86}, 284 (2001).


\bibitem{ref21}
Q.-f. Sun, J. Wang, and T.-h. Lin, Phys. Rev. B {\bf 62}, 648 (2000).

\bibitem{ref22}
Q.-f. Sun, B.-g. Wang, J. Wang, and T.-h. Lin, Phys. Rev. B 
{\bf 61}, 4754 (2000).

\bibitem{ref23}   
N. S. Wingreen, Antti-Pekka Jauho, and Y. Meir,  Phys.
Rev. B{\bf 48}, 8487 (1993).

\bibitem{ref24}   
Antti-Pekka Jauho, N. S. Wingreen, and Y. Meir, Phys. Rev. B 
{\bf 50},  5528 (1994).

\bibitem{ref25}
If the inelastic processes inside the superconducting lead are considered,
$\epsilon$ will have a small positive imaginary part $\delta$, then 
$\beta_L(\epsilon+i\delta) =
\frac{\epsilon+i\delta}{i\sqrt{\Delta_L^2-(\epsilon+i\delta)^2}}$
for $\Delta_L>|\epsilon|$, and 
$\beta_L(\epsilon+i\delta) = 
\frac{(\epsilon+i\delta)Sign(\epsilon)}
{\sqrt{(\epsilon+i\delta)^2-\Delta_L^2}}$ for $\Delta_L<|\epsilon|$.

\bibitem{ref26}
Q.-f. Sun, J. Wang, and T.-h. Lin, Phys. Rev. B {\bf 59}, 13126 (1999).


\bibitem{ref27}
J. Wang, Y. Wei, H. Guo, Q.-f. Sun, and T.-h. Lin, 
will be published in Phys. Rev. B

\bibitem{ref28}
In this expression, the condition of $V_L=0$ has been used. If $V_L\not=0$,
$G_{12}(t,t_1)$ may be not equal to $G_{12}(t+T,t_1+T)$.

\bibitem{ref29}
Even if the intradot level $\epsilon_d$ and the hopping elements $t_{L(R)}$
are dependent on spin index $\sigma$, the equation
$\hat{G}^<(t,t_1)=-[\hat{G}^<(t_1,t)]^{\dagger}$ is still valid. 

\bibitem{ref30}
Strictly speaking, the time-averaged distribution of electrons should be 
calculated as follows: first one calculates the ratio of time-dependent 
occupation number and the LDOS; second one performs the time-average
of this ratio. However, due to the fact that high-order components of the 
distribution are generally small,
(except at small bias for those devices with large $\Gamma$), 
the distribution can be well approximated 
by the ratio of average quantities.

\bibitem{ref31}
H. Pothier, S. Gueron, N.O. Birge, D. Esteve, and M.H. Devoret, Phys. Rev. Lett.
{\bf 79}, 3490 (1997).

\bibitem{ref32}
F. Pierre, A. Anthore, H. Pothier, C. Urbina, and D. Esteve, Phys. Rev. Lett.
{\bf 86}, 1078 (2001).

\bibitem{ref33}
G.E. Blonder, M. Tinkham, and T.M. Klapwijk, Phys. Rev. B {\bf 25}, 4515 (1982);
M. Octavio, M. Tinkham, G.E. Blonder, and T.M. Klapwijk,
Phys. Rev. B {\bf 27}, 6739 (1983).

\bibitem{ref34}
A.W. Kleinsasser, R.E. Miller, W.H. Mallison, and G.B. Arnold, 
Phys. Rev. Lett. {\bf 72}, 1738 (1994); 
N. van der Post, E.T. Peters, I.K. Yanson, and J.M. van Ruitenbeek,
Phys. Rev. Lett. {\bf 73}, 2611 (1994).

\end{references}
\end{document}